\documentclass[preprint]{aastex}

\usepackage{amsmath,amssymb,graphicx}

\shorttitle{Dynamo Waves at high $Rm$}
\shortauthors{Cattaneo and Tobias}

\begin{document}


\title{On Large-Scale Dynamo Action at High Magnetic Reynolds Number}


\author{F. Cattaneo}
\affil{Department of Astronomy and Astrophysics and The Computation Institute, University of Chicago,
5735 South Ellis Avenue, Chicago IL 60637}

\and

\author{S.M. Tobias}
\affil{Department of Applied Mathematics, University of Leeds, Leeds,
  LS2 9JT, U.K.}
\email{smt@maths.leeds.ac.uk}




\begin{abstract}
We consider the generation of magnetic activity  --- dynamo waves --- in the astrophysical limit of 
very large magnetic Reynolds number. We consider kinematic dynamo action for a system consisting of helical
flow and large-scale shear. We demonstrate that large-scale dynamo waves persist at high $Rm$ if the
helical flow is characterised by a narrow band of spatial scales and the shear is large enough.  However for a wide band 
of scales the dynamo becomes small-scale with a further increase of $Rm$, with dynamo waves re-emerging only if
the shear is then increased. We show that at high $Rm$ the key effect of the shear is to suppress small-scale dynamo action, allowing large-scale dynamo action to be observed. 
We conjecture that this supports a general ``suppression principle" --- large-scale dynamo action can only be observed if there is a mechanism that suppresses the small-scale fluctuations.
\end{abstract}





\section{Introduction}

\label{intro}

The emergence of organized, systematic astrophysical magnetic fields from a turbulent environment remains
one of the outstanding conundrums of astrophysics \citep{Parker:79}. The magnetic fields of stars and galaxies often display organization on timescales and lengthscales large compared with that of the underlying turbulence. The classical example is the eleven year solar activity cycle where the emergence of active regions at the solar surface is the manifestation of a magnetic field generated deep within the Sun with a temporal coherence 
much longer than the flows in the convection zone and a global spatial coherence \citep{tw:2007,Weiss-Thompson:09}.

It is now widely accepted that such astrophysical magnetic fields are generated by hydromagnetic dynamos. The first step towards an astrophysical theory of large-scale field generation was taken by Parker in an
epoch-making paper \citep{Parker:1955}. He pointed out that significant progress could be made by deriving equations governing the evolution of the average of the magnetic field. He demonstrated that for cyclonic convection (convection lacking reflectional symmetry) the net effect of small-scale interactions was a mean electromotive force with a component aligned with the mean magnetic field, that was capable of regenerating large-scale magnetic fields. In this paradigm the interactions at the small-scales were parameterised using transport coefficients. Moreover he demonstrated that in the presence of a large-scale differential rotation or shear, the solution of the averaged equation has the form of a travelling dynamo wave. This approach was independently formalised as mean-field electrodynamics by the Potsdam school \citep{skr:1966,kr:1980}. They showed that the mean induction could be non-zero only for flows lacking reflectional symmetry \citep{moff:78}. It turns out that a very convenient measure of the lack of reflectional symmetry in a fluid flow is the kinetic helicity, which is a measure of the degree of correlation of the flow's velocity and vorticity.
These considerations led to the general belief that differential rotation and helical turbulence were the two building blocks necessary to give rise to cyclic magnetic activity. This belief was further reinforced by the fact that helical flows, like for example the cyclonic events of Parker, arise naturally in rotating stratified turbulence.

The mathematical complexities in calculating the average induction from the properties of the underlying turbulence, as measured by say the $\alpha$-effect, are such that at the moment, it can only be carried out in two limiting cases --- small magnetic Reynolds numbers ($Rm$) or short correlation time turbulence. In most astrophysical circumstances neither of these is valid; correlation times are comparable with the turnover times in the turbulence and the magnetic Reynolds numbers are huge. In general it is assumed that even for finite correlation times and for large $Rm$ these basic mechanisms remain valid. Therefore this idea of helical turbulence and large-scale shear has become the
intellectual framework used to understand cyclic magnetic activity. It is thus important to develop systems in which we can demonstrate that these assumptions are correct.

The limit of large magnetic Reynolds number is extremely delicate and it is useful to identify two different issues. The first of these is whether for a given flow, dynamo action can be sustained at all in this limit. This is the well-known fast dynamo problem \citep{cg:1995}. It is an intriguing property of dynamos that on the one hand dynamo action succeeds if inductive processes overcome diffusion, on the other hand a dynamo needs diffusion to operate and bring about the necessary changes in magnetic topology.  The second issue relates to the averaging process. At high $Rm$ often what happens is that small-scale fluctuations become dominant and it becomes difficult to extract meaningful averages
from the fluctuations \citep{ch:2009}. This problem manifests itself in a divergence between the solution of the averaged equations and the averages of the solutions of the exact equations \citep{bcr:2005}. Recently some progress was made by \citet{tc:2013} - hereinafter TC13 - in constructing a system 
consisting of large-scale shear and helical flows that yielded demonstrable large-scale cyclic behaviour even at large $Rm$. In this paper we build upon this model and extend to consider in detail the inductive mechanisms that allow cyclic behaviour to emerge, and we discuss more complicated cases in which flows on a large range of scales are present. 

\section{Theoretical Considerations}
\label{theory}

In this section we wish to consider general properties of systems that consist of a helical flow and a shear. Helical flows are necessary for mean induction; it is therefore obvious that we should include them in our model. The role of the shear is more subtle and so deserves closer attention.\footnote{Notice that these statements would have been exactly the opposite a few years ago, shear was straightforward and mean induction
was considered subtle.} There are at least four potentially important effects that are associated with the presence of shear. The first of these is that in a homogeneous isotropic system the addition of shear breaks left-right symmetry  and so leads naturally to wave-like solutions being preferred; here we are assuming that the time-averaged helicity is one signed. 
Because the shear is large-scale it survives the averaging process and therefore breaks the symmetry in both the small and large-scale equations.
The second is the well-known effect that shear can enhance  diffusion in the direction transverse to the shear, by increasing the effective wavenumbers of the magnetic field. In certain circumstances this can be an effective way of removing small-scale fluctuations. Another interesting property of shear
is that it can give additional contributions to the mean electromotive force \citep{rk:2007,Yetal:2008,kkp:2010,srisin:2010,kss:2012,hp:2013}. Finally the addition  of shear  can alter the Lagrangian properties of flow. Although in general it is difficult to say precisely what the effect of the shear in Lagrangian trajectories is, for flows that are strongly chaotic the effects of a shear is almost always to reduce the largest Lyapunov exponent, which can lead to a decrease in the dynamo growth-rate.

With these considerations in mind, it is natural to ask which of these becomes important at high $Rm$. In general it depends on the flow. For the 
present discussion it is important to distinguish between flows on one (or a small number of) characteristic scale  and those like turbulence with a whole spectrum of scales; in both cases here we are considering periodic cellular flows. We begin by discussing the (simpler) one scale case. Suppose that we consider a dynamo that consists of a helical flow that is strongly chaotic. Here strongly chaotic means that the largest Lyapunov exponent is comparable with the turnover frequency of the flow, which we believe to be the generic case for turbulence. Flows of these types are often fast dynamos in sense that in the limit of large $Rm$ the dynamo growth-rate is  comparable with the turnover frequency of the flow. Furthermore if the flow has the ``quick dynamo property"\footnote{A dynamo is defined to be ``quick" if it attains a maximum growth-rate quickly, i.e.\ for a magnetic Reynolds number not much larger than that needed for the dynamo to activate}, the approach to this asymptotic growth-rate will occur at moderate Reynolds numbers \citep{tc:2008} and if the flow has the fast dynamo property it will persist even at extremely high $Rm$.
In a homogeneous system the magnetic field generated by the flow will have thin current sheets whose thickness is controlled by $Rm$ (scaling as $Rm^{-1/2}$) and radius of curvature comparable with the scale of variation of the flow. The resulting magnetic field will have a periodicity comparable with the driving flow. Adding a small amount of shear to this cellular flow is will have two effects. The first is that the growth-rate of the dynamo will decrease slightly \citep{ct:2005}. This is because we have assumed that the original flow was strongly chaotic ---  adding a steady flow that leads to the at most algebraic divergence of neighbouring trajectories will therefore lead to a reduction of the dynamo growth-rate. The other effect is that the addition of shear will bring in exponentially growing dynamo wave solutions with a growth-rate small compared with that of the fastest growing mode. In an initial value problem these modes will therefore not be seen at small values of the shear. If one increases the strength of the shear these effects will continue; the growth-rate of the fastest-growing mode will reduce, that of the dynamo wave solution will increase until the dynamo wave will manifest itself in an initial value problem. This most likely occurs when the shear-rate across a cell is comparable with the turnover frequency. If $Rm$ is now increased further what happens to the structure of the solution? Because the fast dynamo property and the quick dynamo property depend only on the Lagrangian structure of the flow; providing these are not affected by the structure of the shear, then increasing $Rm$ should not affect the dynamo growth-rate. Other structural properties of the solution will change, for example the thickness of current sheets and the ratio of unsigned to signed flux, but the growth-rate, frequency and the large-scale spatial structure of the solution should remain the same.

This behaviour should be contrasted with the more complicated case where one has a large range of spatial scales.
We begin again by considering the case with no shear. In a typical turbulent realisation there is a power law spectrum for the energy and a corresponding one for the relative helicity. If the slope of the spectrum is sufficiently shallow,  the energy decreases with increasing scale, whilst the shear rate increases. Therefore, the largest eddies have the largest amplitude, and therefore the highest (scale-dependent) $Rm(k)$, whilst the smallest eddies have the fastest (scale-dependent) turnover time $T(k)$. This is the case for example for Kolmogorov turbulence, with a spectral exponent of $-5/3$. It was argued in Cattaneo \& Tobias (2008) that for this configuration, the growth-rate is determined by the eddies for which $Rm(k) \sim Rm_q$ (where $Rm_q$ is the value of $Rm$ for which the dynamo reaches its asymptotic value) and will be comparable with the turnover frequency of the eddies at that scale. The value of $Rm_q$  depends on the geometry of the dynamo eddies, but for quick dynamos $Rm_q \sim 10-50$. We term this scale that controls the dynamo properties of the flow the ``dynamo scale".  Notice that if the overall $Rm$ is increased, say by reducing the magnetic diffusivity, then the ``dynamo scale" will move to higher wavenumbers (smaller scales). We now consider the case where a small amount of shear is added to this flow with a large range of scales. The ability of the shear to deform the eddies will again be scale-dependent, with the shear more easily deforming the large-scale eddies than the small-scale eddies. Thus the immediate effect is to have dynamo wave solutions associated with a band of small wavenumbers, but the growth-rate of these dynamo waves will be small compared with the growth-rate of the dynamo associated with the dynamo scale. If the amplitude of the shear is increased, the high-wavenumber end of this shear-affected (or dynamo wave) band will slide towards higher wavenumbers, until eventually it will coincide with the dynamo scale. For this value of the shear the fastest growing mode is a dynamo wave, with a growth-rate that will most likely be lower than the original unsheared dynamo. How does this system respond to further changes in overall $Rm$? If $Rm$ is decreased the dynamo scale moves to smaller wavenumbers with smaller growthrates whilst remaining a dynamo wave. However if $Rm$ is increased the dynamo scale moves to larger wavenumbers that are outside the dynamo wave band. The fastest growing mode will revert back to a small-scale dynamo with a correspondingly higher growth-rate. 

Of course in a realistic astrophysical flow, the helicity will be a function of scale, with scales below the Rossby Radius of deformation having correspondingly less helicity. We shall consider this modification in the Discussion section.

\section{Formulation of the Model}
\label{formulation}

While the considerations in the previous section were somewhat general and abstract, it is beneficial to illustrate some of this behaviour with a concrete example.  Our basic building block for the velocity is a cellular flow, with a well-defined characteristic scale and turnover time, well-established dynamo properties, and such that it can be realised at high $Rm$ with reasonable computational resources. One very natural choice
is therefore to consider the circularly polarised incompressible Galloway-Proctor flow at scale $k$ \citep{ct:2005}. Here we are considering Cartesian co-ordinates $(x,y,z)$ on a $2 \pi$-periodic domain.
This flow takes the form
\begin{equation}
{\bf u}_k=A_k \left( \partial_y \psi_k, -\partial_x \psi_k, k \psi_k \right),
\end{equation}
where
\begin{equation}
\psi_k(x,y,t) = (\sin k ((x-\xi_k)+\cos \omega_k t)
+\cos k ((y-\eta_k)+ \sin \omega_k t)).
\end{equation}
Flows of this type are called 2.5-dimensional as they have all three components but only depend on two spatial dimensions. 
This flow is also maximally helical --- it takes the form of an infinite array of clockwise and anti-clockwise rotating helices
such that the origin of the pattern itself rotates in a circle with frequency $\omega_k$. Here $A_k$ is an amplitude and $\xi_k$ and $\eta_k$ are 
offsets that can be varied so as to decorrelate the pattern. Here they are random constants that are reset every $\tau_d$, which can therefore be regarded as a decorrelation time. 

The dynamo properties of this flow are well-understood. Because the velocity does not depend on the $z$ co-ordinate, the kinematic dynamo problem is separable and magnetic fields of the form
\begin{equation}
{\bf B}(x,y,z,t) = {\bf b}(x,y,t) \exp i k_z z
\end{equation}
can be sought. If the decorrelation time were infinite then the velocity is exactly time-periodic and the magnetic field
has a well-defined growth-rate. If the decorrelation time is finite then the solutions have a well-defined average
growth rate (averaged over times longer than the decorrelation time). In general the growth-rate depends on $k_z$ which can be
regarded as a parameter in the problem. The dynamo properties therefore depend on $k$, $k_z$, $A_k$ and $\omega_k$ --- if all of these are order unity then dynamo sets in when $Rm$ is order one, the $k_z$ of maximum growth-rate is of order one and the maximum growth-rate as $Rm$ gets large is itself of order one. Furthermore the actual growth-rate is a few percent away from the maximal value by $Rm \sim 10$. Thus these flows have both the fast and quick dynamo property. The corresponding eigenfunctions for the magnetic field have the same horizontal periodicity as the basic flow and take the form of a helical wave with a spatial period of order one. The horizontal average of this eigenfunction has the form of a uniform horizontal field whose direction rotates by $2 \pi$ over an inverse lengthscale $k_z$. The polarization of the helical wave (i.e.\ whether the average rotates clockwise or anticlockwise) depends on the sign of the helicity.

In general we wish to consider a superposition of these flows. Therefore at each scale $k$ we are required to set $A_k$ and $\omega_k$. We
are free to choose $A_k$ to mimic the properties of any spectrum of turbulence. Having chosen $A_k$ there is then a unique choice of $\omega_k $ such that the associated dynamo action at scale $k$ has the same asymptotic growth-rate measured in units of the local turnover time. This is given by $\omega_k \sim k^2 A_k$  --- for this choice the ratio of asymptotic growth-rate to turnover frequency is independent of $k$ \citep[see e.g.][]{ct:2005}. Therefore at their own scale, all of these dynamos look the same.

With all this machinery behind us, our cellular flow takes the form of a superposition of these flows on scales between $k_{min}$ and $k_{max}$, i.e.\ we set
\begin{equation}
{\bf u}_c= \sum_{k_{min}}^{k_{max}} {\bf u}_k.
\end{equation}
 We choose  $A(k)=k^{-\beta}$ (with $\beta=4/3$) so that $\omega_k=k^{2-\beta}$
and the decorrelation time $\tau_d = \tau_0 k^{\beta-2}$. With
these scalings  the turnover time $\tau_k$ and the
magnetic Reynolds number $Rm_k$ are themselves functions of $k$ 
given by \citep{tc:2008}
\begin{eqnarray}
\tau_k &\sim& k^{\beta-2}, \\
R_m(k) &\sim& k^{-\beta}.
\label{more-scaling}
\end{eqnarray}
Thus for positive values of $\beta$, $R_m$ decreases with $k$, and for
$\beta <2$ so does the turnover time.  

To this flow we add a steady unidirectional large-scale shear of the form
\begin{equation}
{\bf u}_s = \left(V_0 \sin y, 0, 0\right).
\end{equation}

For this prescribed flow ${\bf u}= {\bf u}_s+{\bf u}_c$ we solve the induction equation 
for ${\bf b}$ as an initial value problem for a given $k_z$ using a pseudospectral code, with typical resolutions of
$2048^2$. 
For random initial conditions this method picks up the fastest growing eigenfunction within a few
turnovers.

\section{Results}

\subsection{A narrow band velocity}

We begin this section by considering cases with a narrow band of scales --- thus we set $k_{min}=8$ and $k_{max}=12$, and examine in detail
two cases, one with the decorrelation time comparable with the turnover time ($\tau_0=2.0$) and the other with a short decorrelation time ($\tau_0=0.1$). Initially we set
$\tau_0=2.0$.
Figure~1 shows contours of the  streamfunction of the total horizontal velocity for two cases, one unsheared and one with a strong shear. 
In the first case the flow is entirely cellular and all streamlines are closed. In the second, on the other hand, the streamlines in the neighbourhood of the highest velocity
traverse the flow and lead to the formation of open channels.
The first process is now to calculate the optimum value of $k_z$ for each value of the shear. Figure~2 shows typical growth-rate curves
as a function of $k_z$ with differing $V_0$. It is clear from this figure that for each value of $V_0$ chosen there is a well-defined $k_z$ that yields
a maximum in the growth-rate. The value of $k_z$ that yields this maximum may also depend on the value of $V_0$. We note here that for moderate $k_z$ the maximum
of the curve is pretty flat, so precise optimisation of the wavenumber is not required. Of course, increasing $Rm$ leaves the growth-rate at the maximum unchanged
but brings up the growth-rates on the high wavenumber tail of the curve \citep[see also][]{gp:1992}.
From now on, unless otherwise stated, we shall focus on the optimised growth-rate.

Figure~3 shows the effects of increasing the shear. In Figure~3a  the growth-rate is given as a function of $V_0$. For this dynamo, which in the absence
of shear is ``as good as it gets" in the sense that it has reached its optimal growth-rate and is operating in the asymptotic regime, adding the shear reduces the growth-rate. 
As discussed above, the addition of shear has introduced channels to the flow; it has increased the integrability and reduced the chaos
leading to a reduced growth-rate at this high $Rm$. Notice that the shear has its most significant impact for small shears; even a small amount of shear can lead to a diminution of the growth-rate. Once the chaos in the flow has been reduced, subsequent addition of shear has relatively little effect on the dynamo growth-rate.

The most striking effect of the shear is however in the nature of the solutions as shown in Figures~4 and 5. Figure~4a is a 
density plot of $B_x$ in the plane $z=0$. The magnetic field takes the form of small-scale filaments with a thickness that is controlled by the
diffusion which are organised on the same scale as the velocity. There is very little organisation in large scales (either spatial or temporal) as can be seen when the solution is averaged over $x$ and displayed as a function of $y$ and $t$ as in Figure~4b. We contrast this behaviour with that when a shear is present. The snapshot of $B_x$ demonstrates that the magnetic activity is suppressed everywhere except in the regions of strongest shear. 
There appears to be variations of the field on both the large and small (presumably resistive) scale. One can isolate the large-scale behaviour by averaging again over $x$. The result is given in Figure~5b.  The large-scale behaviour appears in the form of the well-known propagating dynamo waves. These propagate in the $z$-direction with the sense of propagation depending on the product of the helicity and shear --- which justifies our use of the term dynamo waves (see Tobias \& Cattaneo 2013). We stress here again that these waves have emerged as the average of the solutions of the
full induction equation and not as the solutions of an averaged equation, although they do coincide with these (mean field) solutions.
This should be contrasted with the unsheared case above in which mean-field theory also predicts the existence of a large-scale solution, that
however can not be identified by this averaging simply because the small-scale fluctuations swamp everything. These considerations, together with
the fact that the growth-rate decreases with increasing shear lends further support to the idea that in this case, in which the unsheared small-scale dynamo is efficient, the main effect of the shear is to suppress the growth of the fluctuations.

We shall now consider a case where the unsheared flow is not ``as good as it gets" --- not even close. Within our framework,  this can be conveniently achieved by decreasing the decorrelation time, so we set $\tau_0=0.1$. Now the optimised growth-rate for the unsheared case is five times smaller than that for the corresponding moderate decorrelation time case described above. Even at this high $Rm$ the dynamo is very
inefficient (meaning it only amplifies field at a fraction of the turnover rate). Further decrease in the decorrelation time would lead to a corresponding decrease in the growth-rate; in fact asymptotically the growth-rate should decrease linearly with $\tau_0$ as predicted by mean field theory \citep{kr:1980}. Figure~6 shows the behaviour of the growth-rate as a function of $V_0$. Clearly the shear is having the opposite effect here --- the shear favours dynamo action and the growth-rate actually increases with increased shear.
This is in accordance with the expectations from mean-field theory and the results of various other studies at low to moderate $Rm$, and we shall return to this point below. For any finite shear the solution again takes the form of dynamo waves, as can be confirmed by examination of Figure~7.

For this inefficient short correlation dynamo, the addition of the shear has increased the dynamo growth-rate. This is a good place to discuss more general principles of the role of shear in determining the dynamo waves growth-rate. If we embrace the general mean field picture for the generation of dynamo waves as an $\alpha \omega$-dynamo, there are two processes that lead to the growth of large-scale fields. The first is a turbulent electromotive force (which is typically linked to the $\alpha$-effect, though in general other effects related to gradients of the field can also be present) which generates the poloidal field from the toroidal field. The second regenerates the toroidal field by good old-fashioned shearing of the large-scale poloidal field and is known as the $\omega$-effect. Clearly (and uncontroversially) increasing the shear boosts the $\omega$-effect, but in fact it can also effect the turbulence and therefore the turbulent induction. Of course in principle shares in turbulent induction can go down as well as up --- this depends on the nature of the turbulence and on the specific nature of induction. Here the shear, after a very small initial increase with $V_0$ for small $V_0$, acts so as to reduce the turbulent induction as shown in Figure~8a, both of which are in agreement with quasilinear calculations of \citet{lk:2009}. Thus  the increase in the growth-rate in the short-correlation time flow can only be attributed to the $\omega$-effect. The fact that the shear acts to decrease the random nature of the flow, can also be seen in Figure~8b, which shows the distribution of the electromotive force
for different values 0f $V_0$. As $V_0$ increases the distribution becomes very sharp and the fluctuations are suppressed. 

We conclude this subsection by considering the effects of increasing $Rm$ further. For the first case $\tau_0=2.0$ a two-fold increase in $Rm$ made no appreciable change to the dynamo growth-rate or the period of the dynamo waves. For the case with short correlation time the same is true for cases with non-zero shear. However there is an increase in dynamo growth-rate with $Rm$ for the unsheared case. We infer that the unsheared dynamo with the short correlation time has not yet reached its asymptotic growth-rate for this value of $Rm$, although we are fully expectant that eventually it will become independent of $Rm$.

\subsection{A broad range of spatial scales}

For these cases with a range of spatial scales we choose $k_ {min}=4$ and $k_{max}=64$ so there is an appreciable difference between the scale dependent turnover time and $Rm$ at the left and right hand end of the spectrum. Specifically for $\beta=4/3$, $Rm(4) \approx 40 Rm(64)$ and
$\tau(4) \approx 6.4\tau(64)$. The effect of an imposed  shear on an eddy at scale $k$ is proportional to $k^{-1}$ so we expect this effect to be 16 times bigger at scale $k=4$ than $k=64$. We consider a case with fixed $V_0=5$ and $\tau_0=2.0$. 

In accordance with the predictions of section~2, changing $Rm$ has two main effects. The first is that the spatial scale associated with the mode of maximum growth rate moves to smaller scales. This is illustrated in 
Figure~9a which shows the growth-rate as a function of $k_z$ for a range of values of $\eta$. Clearly as $\eta$ is decreased the peak in the curves moves to higher wavenumbers. It is interesting to note that the curves
are beginning to accumulate as $Rm$ is increased. This happens because eventually $Rm$ even at $k=64$ begins to become substantial so the behaviour of the system is entirely controlled by the smallest scales.
Figure~9b shows the corresponding increase in optimal growth-rate. This increase reflects the fact that the controlling scales are those that have the shortest turnover time. If the spectrum went on forever, then both
of these trends would also continue. By increasing $Rm$ one could achieve a growth-rate as high as one wanted on a preferred scale as small as one wanted!
Figure~10 shows the eigenfunctions for the preferred mode at low and high $Rm$. For small $Rm$ the ``dynamo scale" can feel the effects of the shear and the corresponding eigenfunction is large scale. However for larger $Rm$ the ``dynamo scale" moves to higher wavenumbers that are not affected by shear and the small-scale dynamo reasserts itself.


\section{Discussion}

In this paper, we have shown that it is possible to get robust large-scale dynamo action in the form of dynamo waves that persists even at high $Rm$. 
These dynamo waves emerge from the interaction of helical turbulence and large-scale shear, but at large $Rm$ the role of the shear is not to boost the induction; rather it suppresses small-scale dynamo action, so that the growth of the large-scale fields can manifest itself. We feel that this is a general principle --- which we term the suppression principle ---  at high $Rm$  mean-field behaviour is apparent {\em only} if there is a mechanism that kills off everything else. Here the effect of the shear, by introducing a strong integrable component to the flow, is to reduce the chaotic properties that are responsible for the otherwise extremely efficient small-scale dynamo action. This should be contrasted with the behaviour at small to moderate $Rm$ where the small-scale dynamo is near marginal anyway and can be overtaken by just boosting the induction a tad. This was illustrated by our results that show that for efficient dynamo action the effect of the shear was to decrease the dynamo growth-rate whereas for the inefficient dynamo the shear boosted the dynamo growth-rate  --- in this case via a simple $\omega$-effect.
In fact the suppression principle can be used to understand why in the case with a broad range of scales one can get small or large-scale dynamo action depending on the value of $Rm$. For small $Rm$ the dynamo scale (as identified by quick dynamo theory) is relatively large and the shear is capable of suppressing the small-scale dynamo associated with this scale --- the result is dynamo waves. For higher $Rm$ the dynamo scale moves down the spectrum and the small-scale dynamo generates field on such a fast timescale that the shear is unable to suppress it --- the result is small-scale fields.

The suppression principle can be invoked to put constraints on dynamo action in astrophysical objects. If the principle applies and large-scale dynamo action is observed then either there must be a mechanism suppressing the generation of magnetic fields at small scales or the two scales are somehow decoupled. Within the kinematic framework dynamo theory is linear and the solutions are superposable; large-scale dynamo action can only be observed therefore if the small-scale dynamo is suppressed. In the cases discussed here the suppression agent was the shear --- though it is possible to  envisage other mechanisms that may act in this way. It is interesting to note that in a rotating, stratified body it is the same processes that lead to the formation of differential rotation that leads to the generation of large-scale dynamo action --- these processes occur on scales  larger than or comparable with the Rossby radius of deformation; scales smaller than this do not feel the effects of rotation. It is therefore not unreasonable to postulate that with the differential rotation and helical turbulence at similar scales the suppression effect can operate effectively.

However in most astrophysical circumstances dynamo action proceeds in the nonlinear regime. Then one can envisage a nonlinear mechanism that leads to either the suppression of the the small-scale dynamo or their saturation 
at a reasonable amplitude. In the saturated regime, whether or not the large-scale field is observable depends on the ratio of the saturation amplitude for the large and small scales rather than their respective growth-rates; this is itself a contentious issue in dynamo theory and has been for twenty years \citep{vc:1992}.


\begin{figure}
\includegraphics[width=3.5in,height=3.5in]{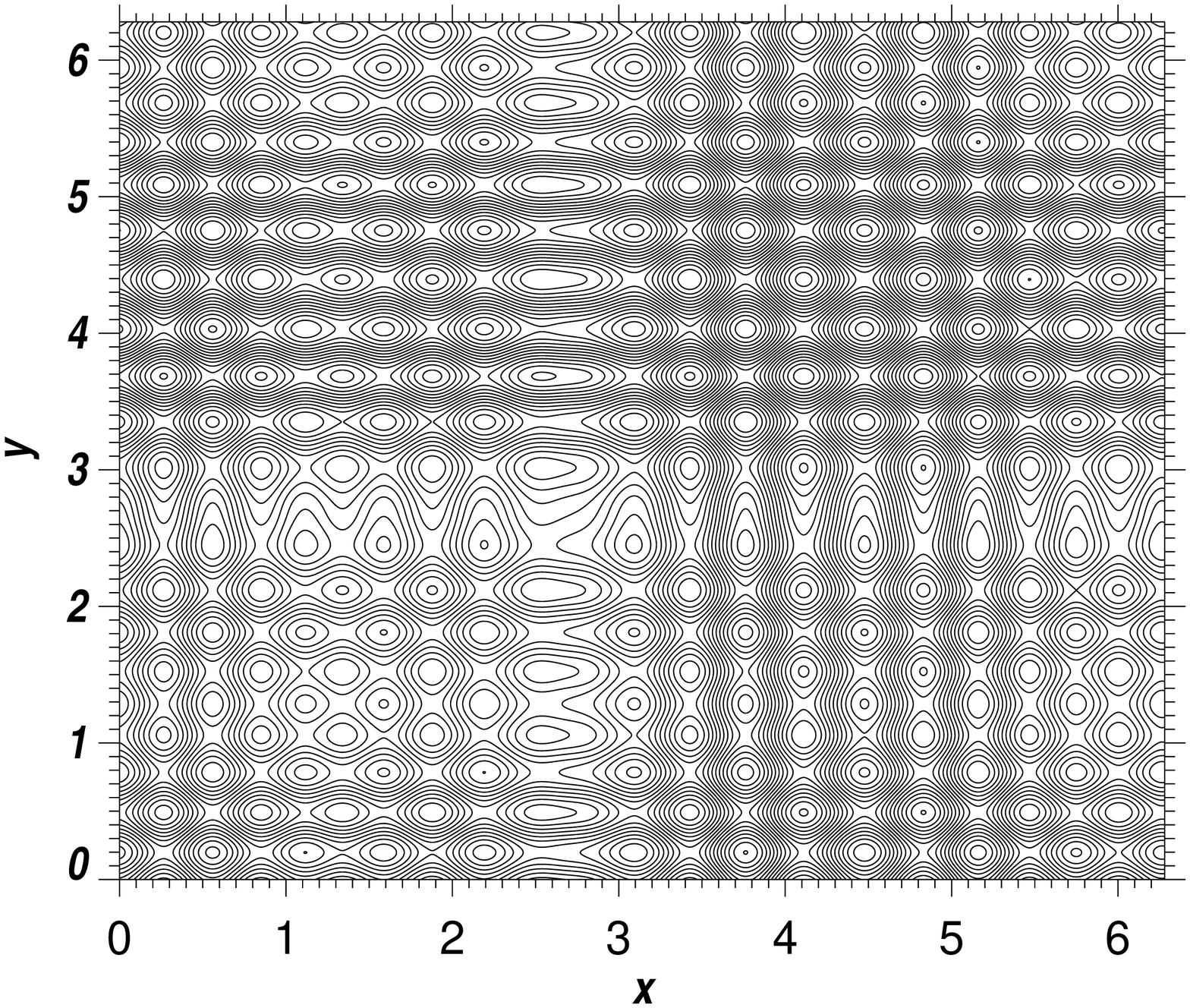}
\includegraphics[width=3.5in,height=3.5in]{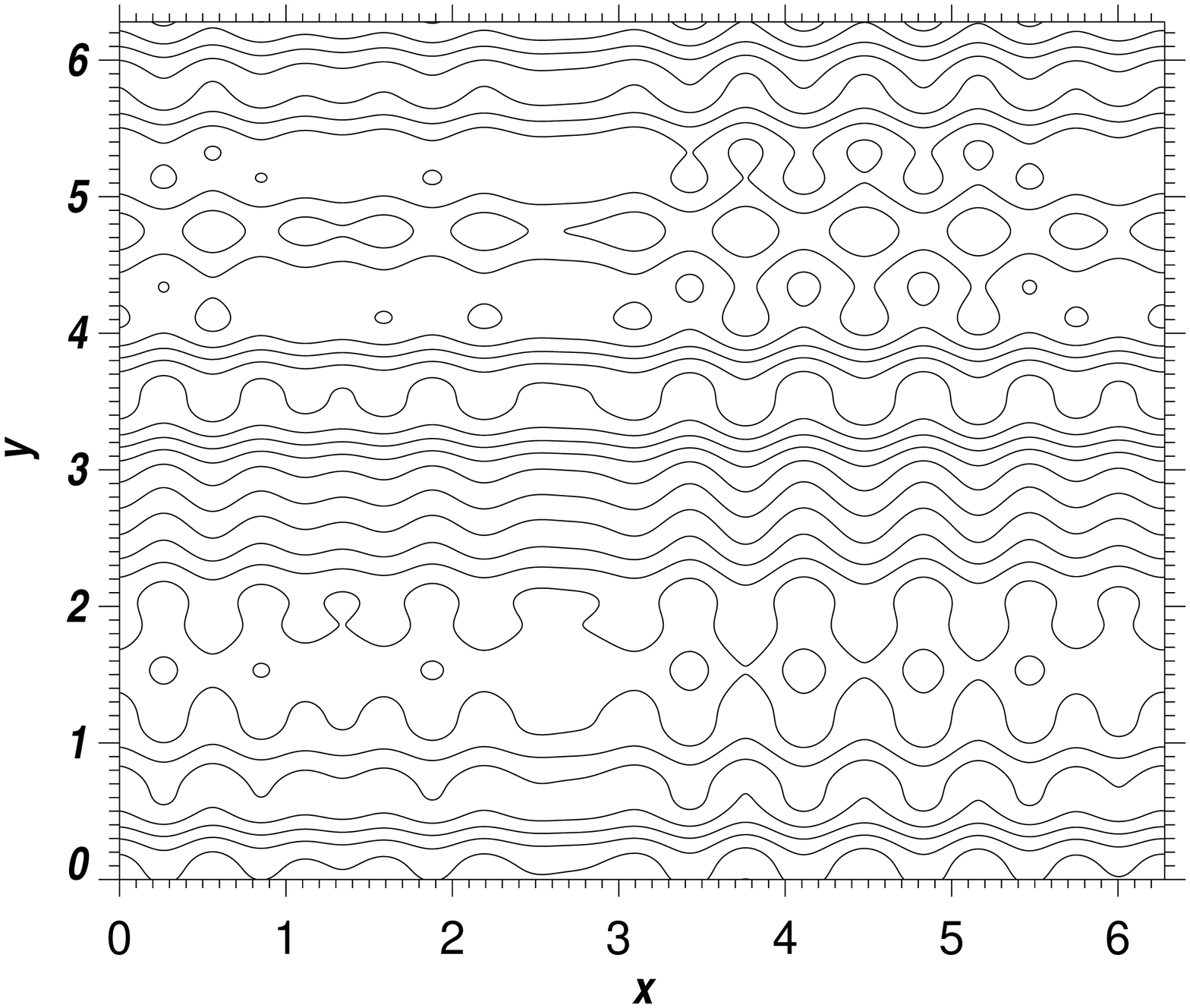}
\caption{Contours of the streamfunction for the total horizontal velocity for (a) $V_0=0$ (b) $V_0=5$.}
\label{figure1}
\end{figure}

\begin{figure}
\centerline{\includegraphics[width=4in,height=4in]{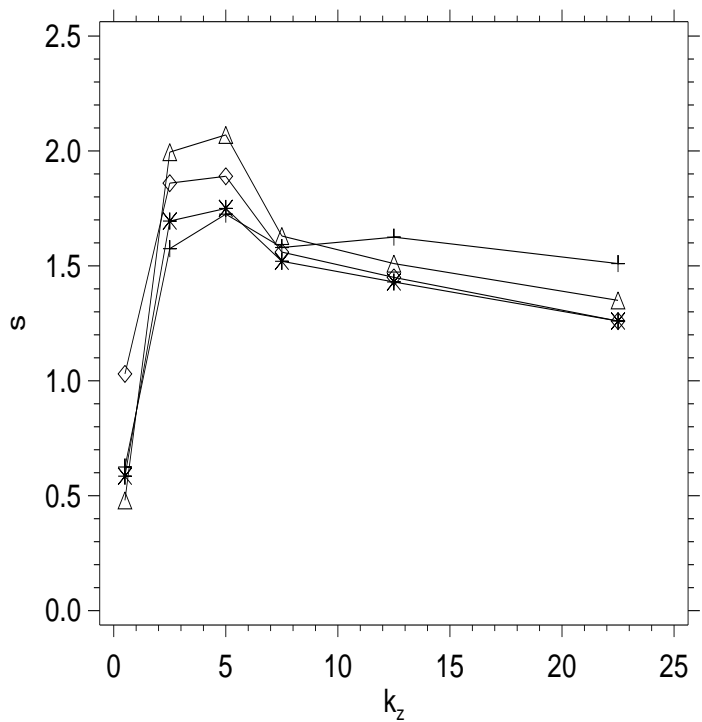}}
\caption{Typical growth-rate curves. Growth-rate as a function of $k_z$ for $V_0=0$ (triangles), $1$ (diamonds) $5$ (asterisks) $10$ (pluses) etc. Here $\eta=10^{-4}$ and
$\tau_0=2.0$}
\label{figure2}
\end{figure}

\begin{figure}
\includegraphics[width=3in,height=3in]{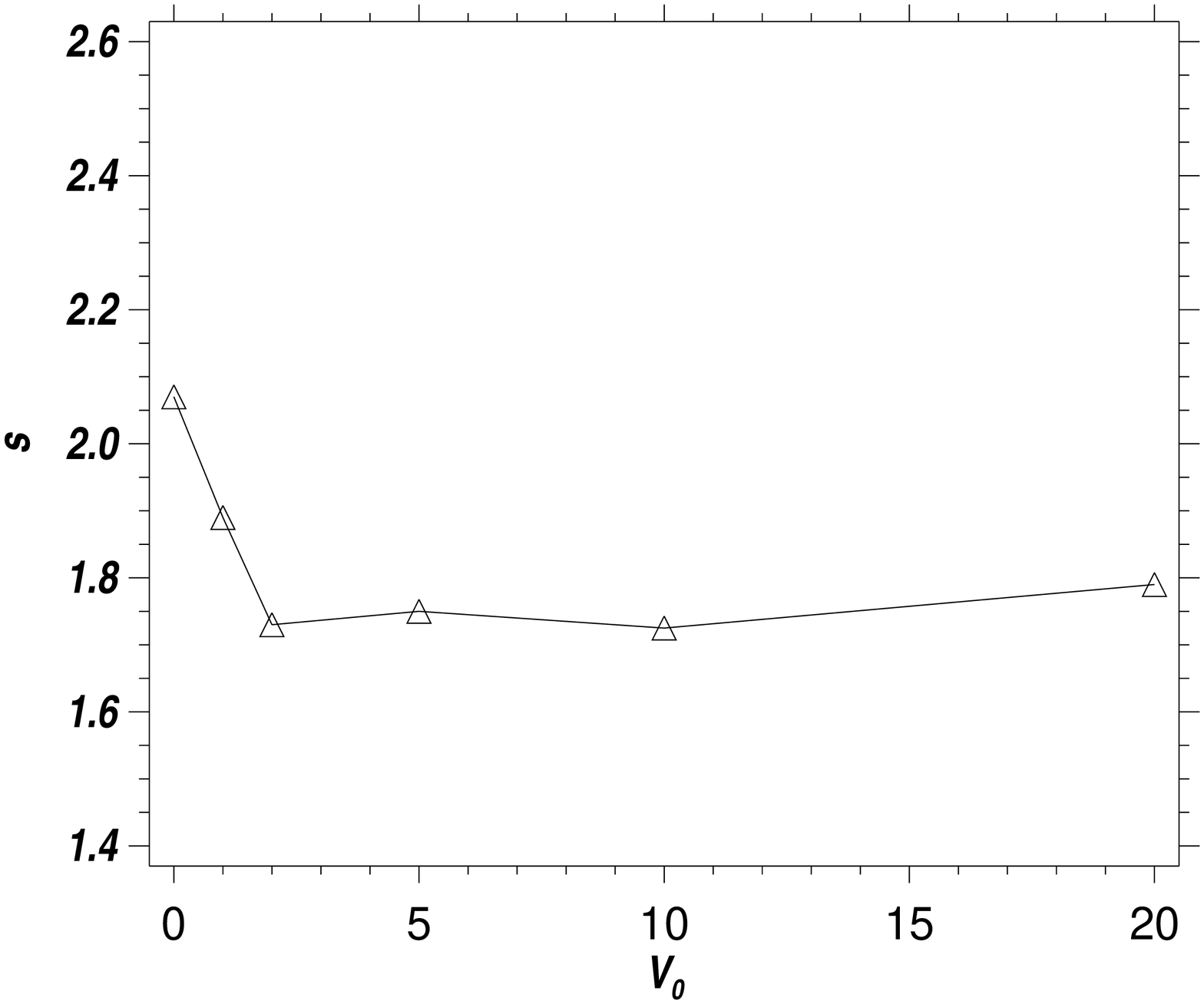}
\includegraphics[width=3in,height=3in]{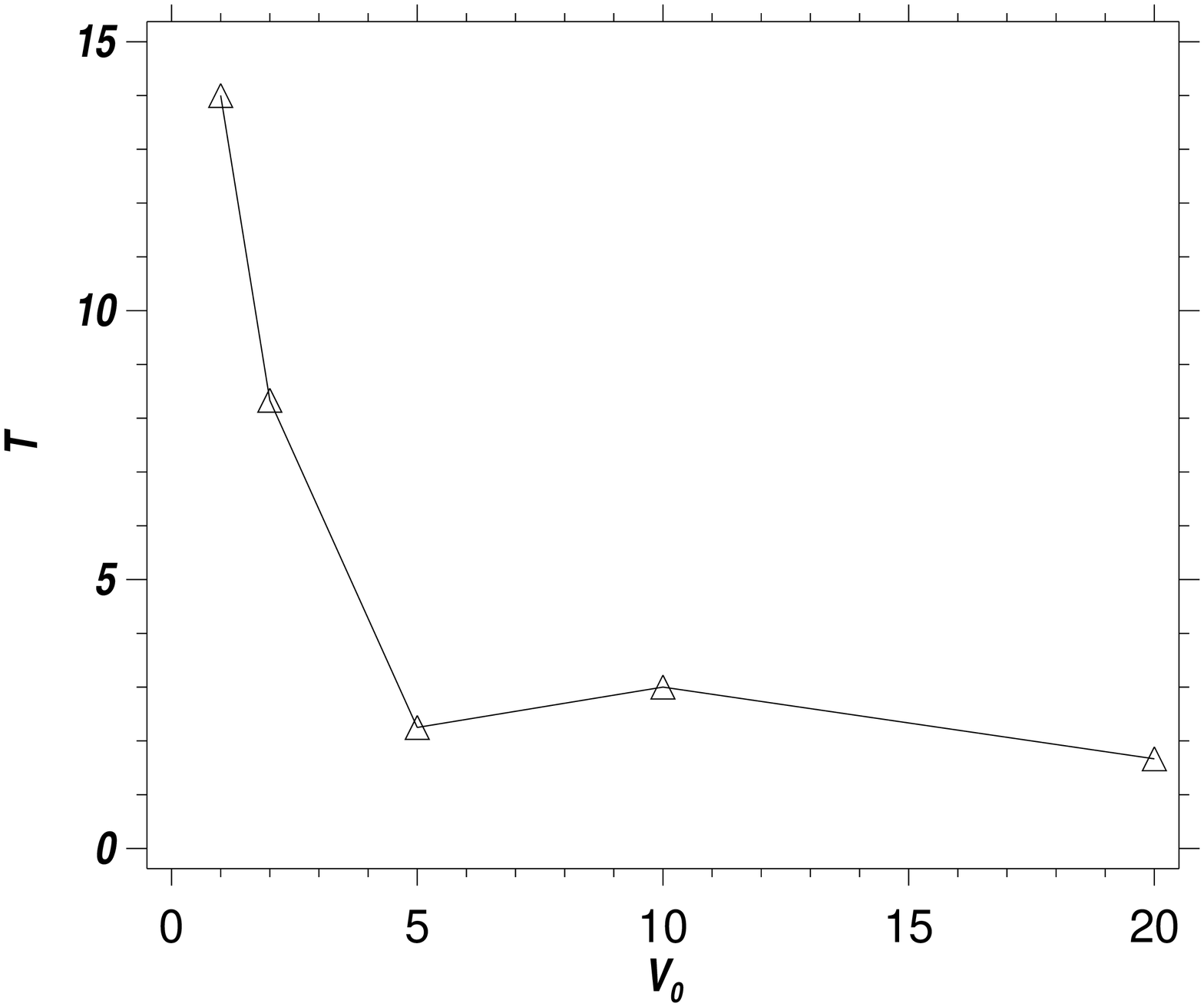}
\caption{(a) Optimal growth-rate and (b) Period of dynamo waves as a function of $V_0$. Other parameters are as for Figure~\ref{figure2}. Note that the growth-rate decreases with
increasing $V_0$}
\label{figure3}
\end{figure}

\begin{figure}
\centerline{\includegraphics[width=3in,height=3in]{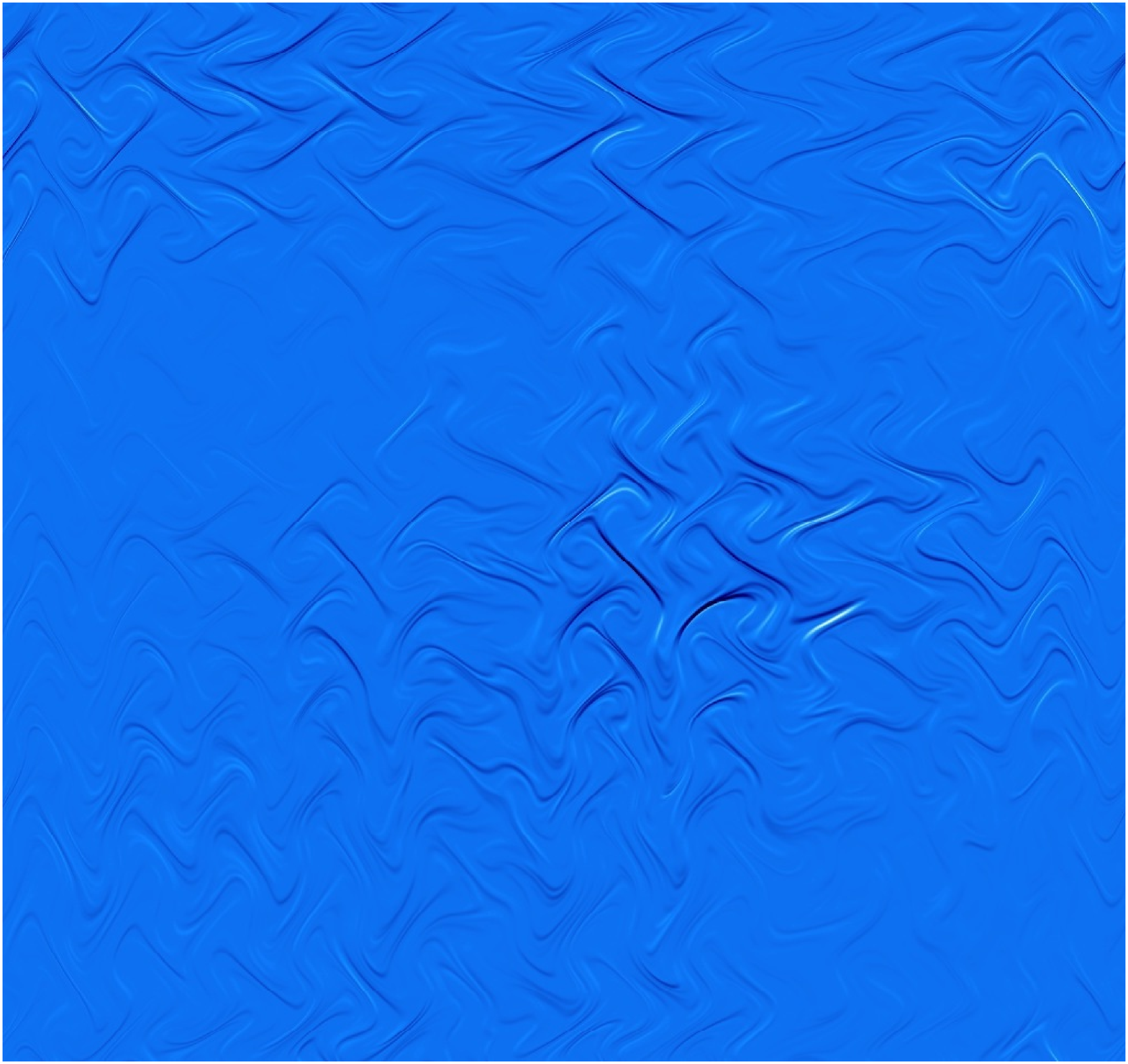}}\hfill\break
\centerline{\includegraphics[width=5in,height=3in]{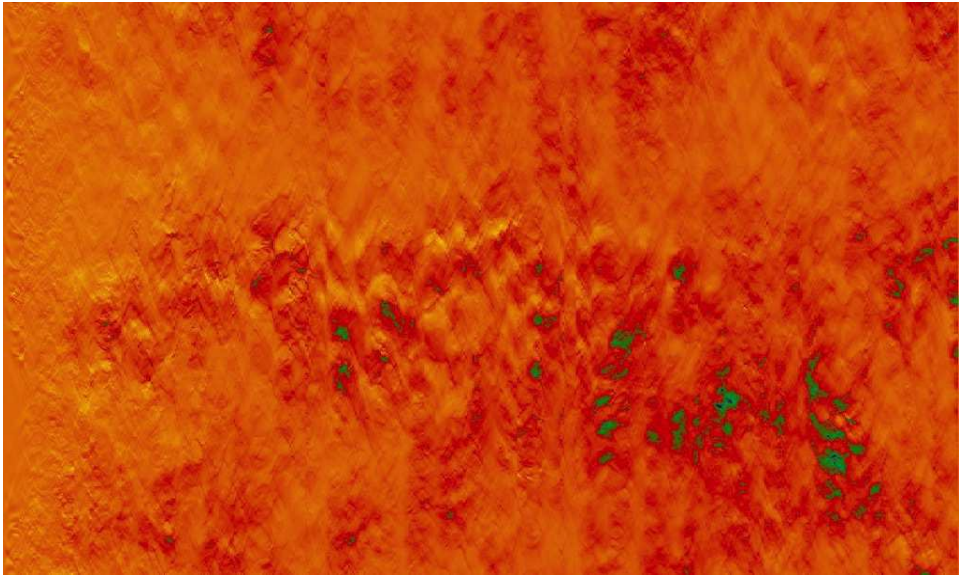}}
\vskip 0.05in
\caption{Small-scale dynamo action for $V_0=0$. (a) Density plot of $B_x(x,y)$ at $z=0$ (b) Space-time plot of the $x$-average of $B_x$ as a function of $y$ and time. No systematic behaviour is visible.}
\label{figure4}
\end{figure}

\begin{figure}
\centerline{\includegraphics[width=3in,height=3in]{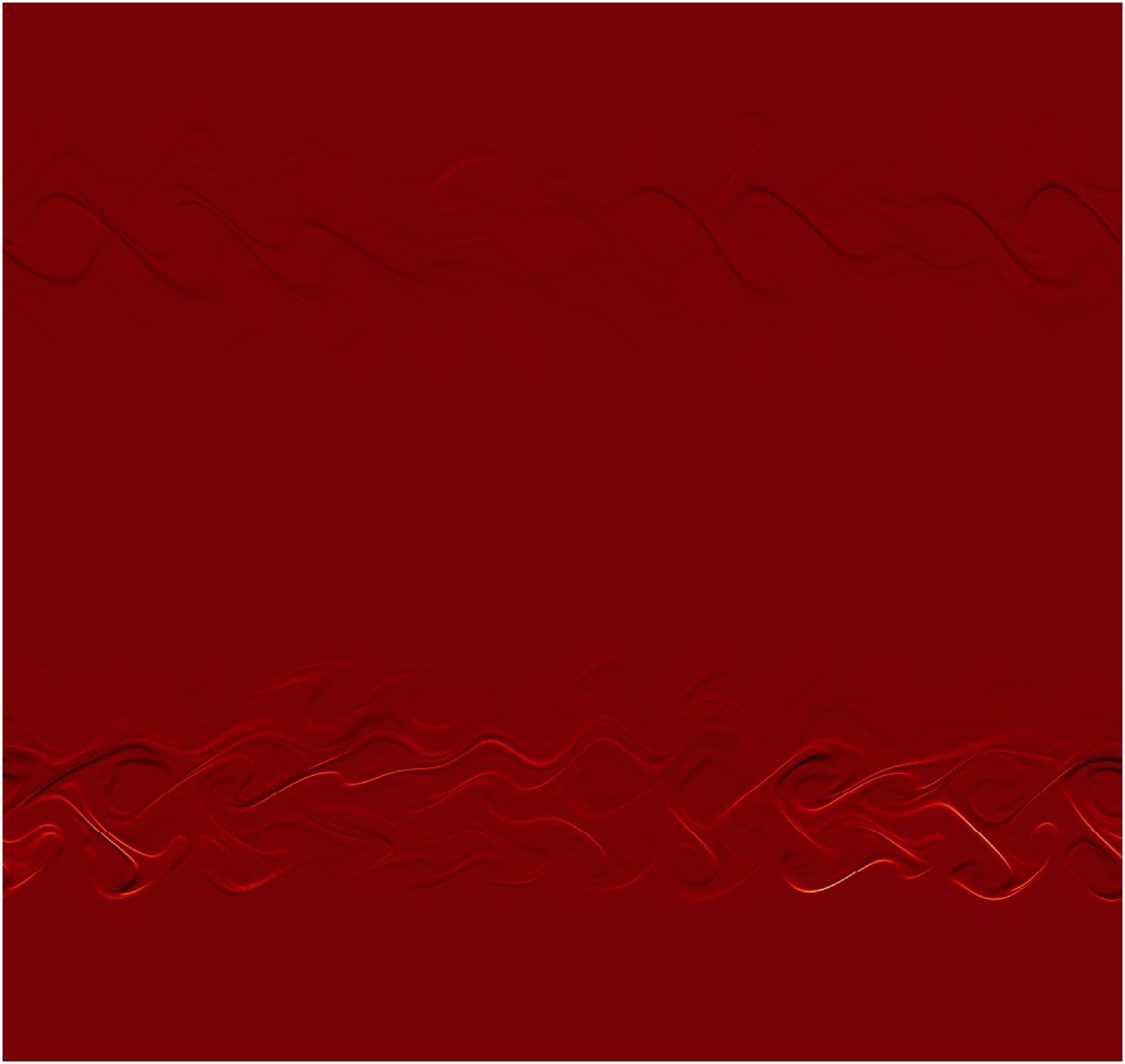}}\hfill\break
\centerline{\includegraphics[width=5in,height=3in]{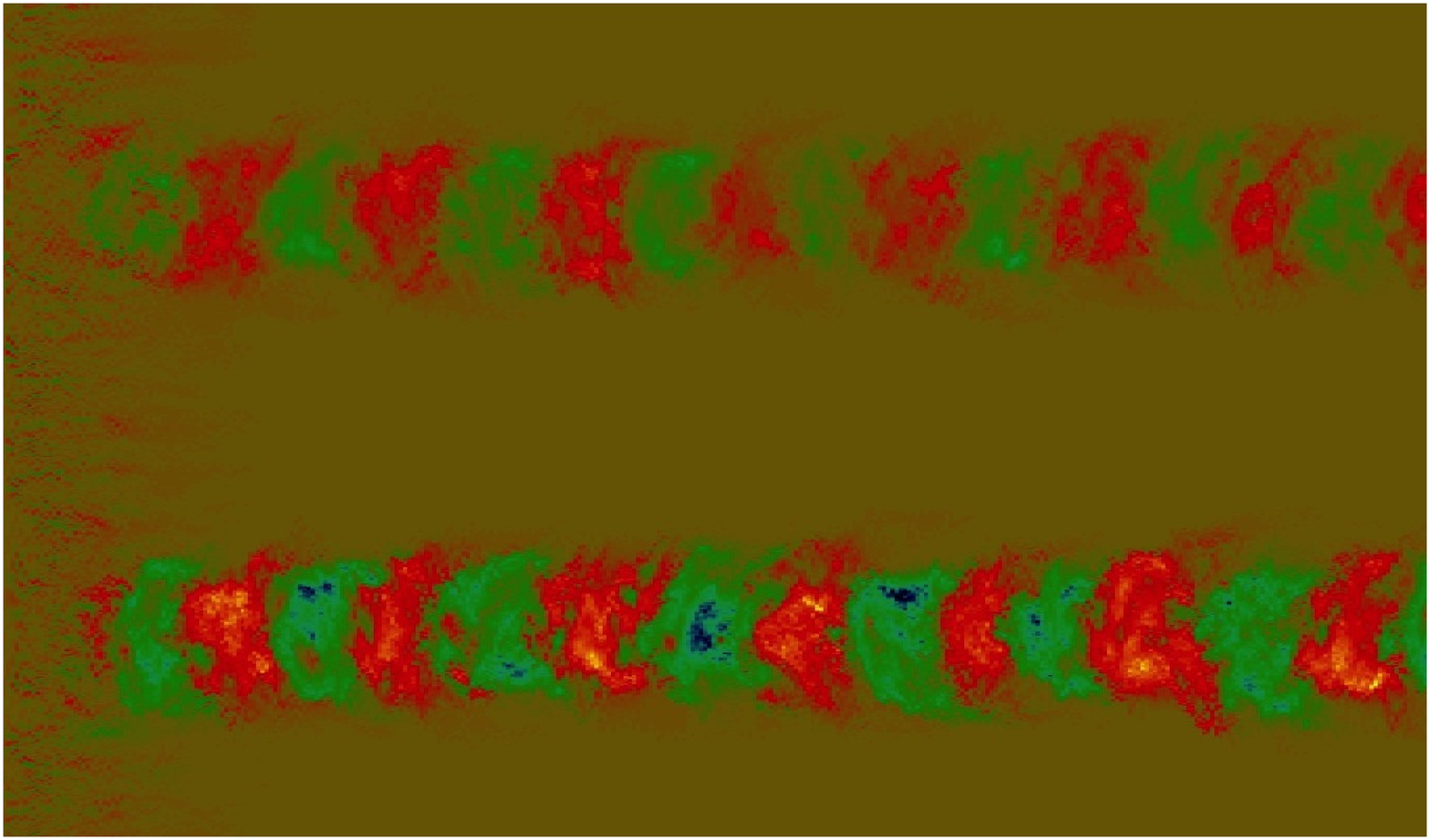}}
\vskip 0.05in
\caption{As for \ref{figure4} but for $V_0=5$. One can now see the dynamo waves clearly.}
\label{figure5}
\end{figure}

\begin{figure}
\centerline{\includegraphics[width=3in,height=3in]{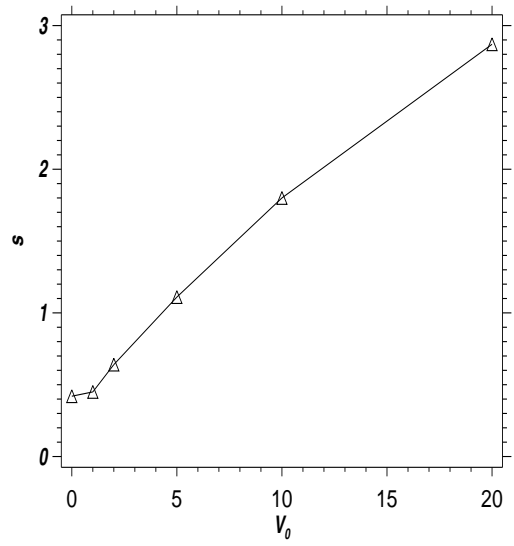}}
\vskip 0.05in
\caption{As for figure~\ref{figure3}, but here $\tau_0=0.1$. Note that the growth-rate increases with
increasing $V_0$ in contrast with Figure~\ref{figure3}.}
\label{figure6}
\end{figure}

\begin{figure}
\centerline{\includegraphics[width=3in,height=3in]{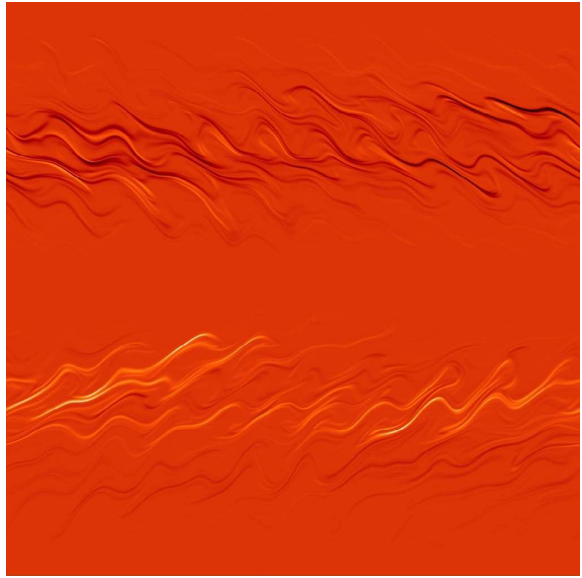}}\hfill\break
\centerline{\includegraphics[width=5in,height=3in]{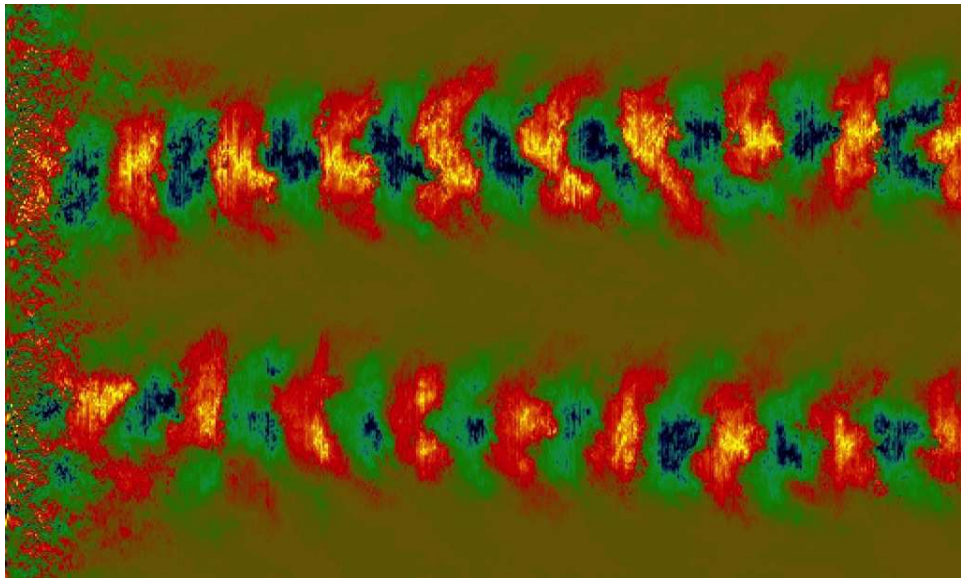}}
\vskip 0.05in
\caption{As for figure~\ref{figure4}, but here $\tau_0=0.1$. Again the dynamo waves are clearly visible.}
\label{figure7}
\end{figure}

\begin{figure}
\includegraphics[width=3in,height=3in]{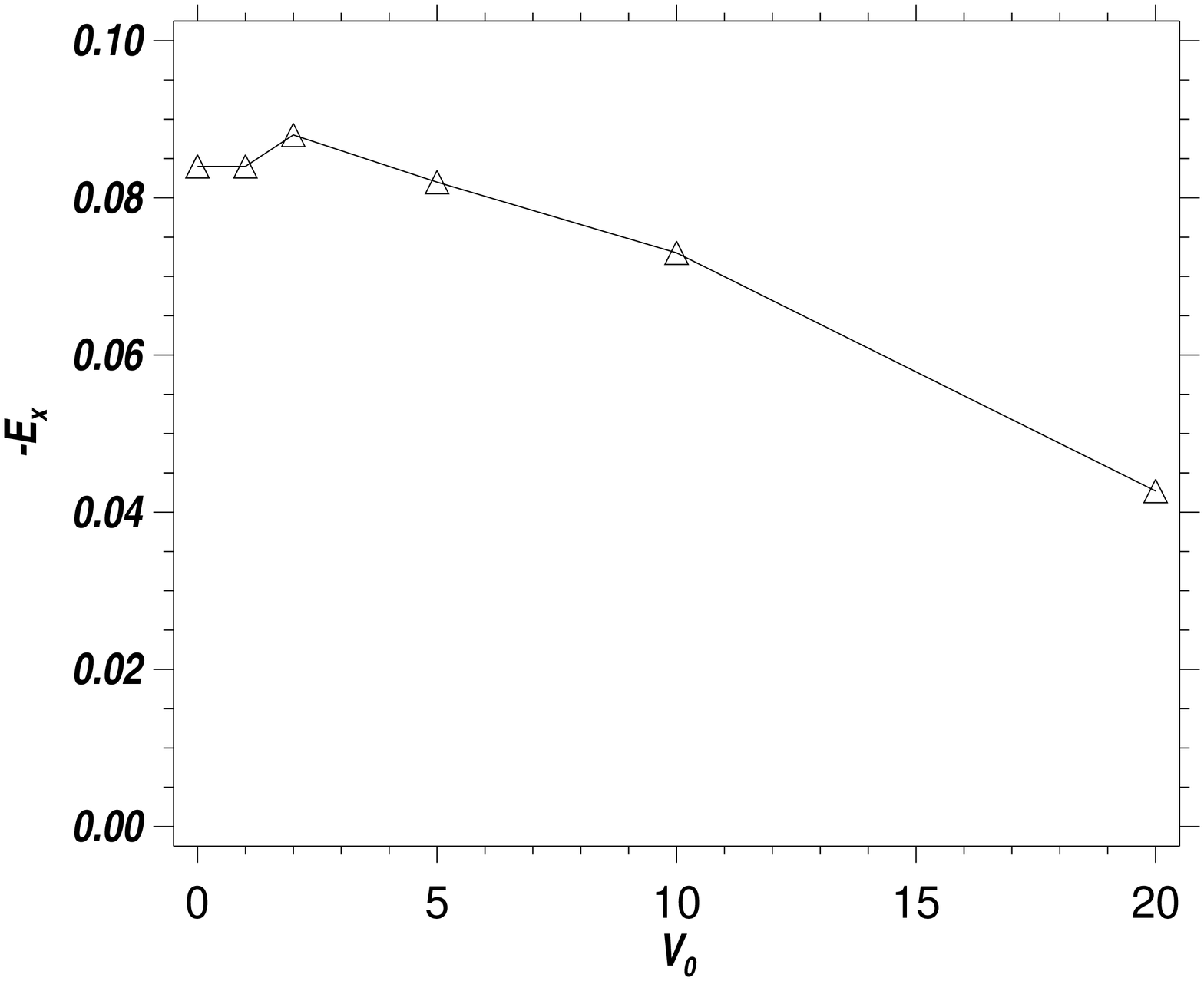}
\includegraphics[width=3in,height=3in]{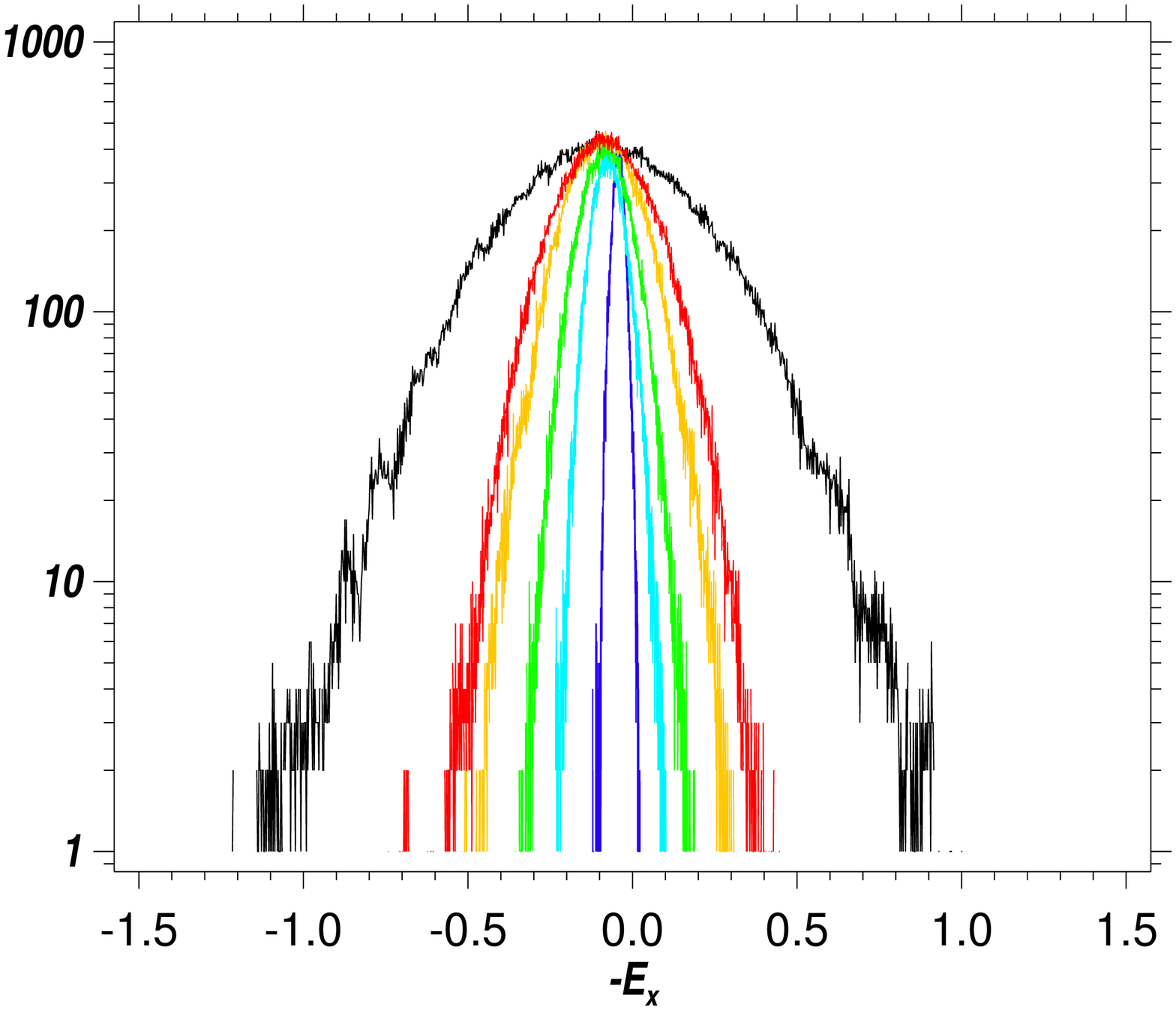}
\vskip 0.05in
\caption{Turbulent electromotive force as a function of $V_0$ for $\tau_0=0.1$. (a) Increase in the shear actually
reduces the average amplitude of the emf (b) Distribution of electromotive force as shear is increased. Here
$V_0=0$ (black) $V_0=1$ (red) $V_0=2$ (yellow) $V_0=5$ (green) $V_0=10$ (cyan) $V_0=20$ (blue).}
\label{figure8}
\end{figure}

\begin{figure}
\includegraphics[width=3in,height=3in]{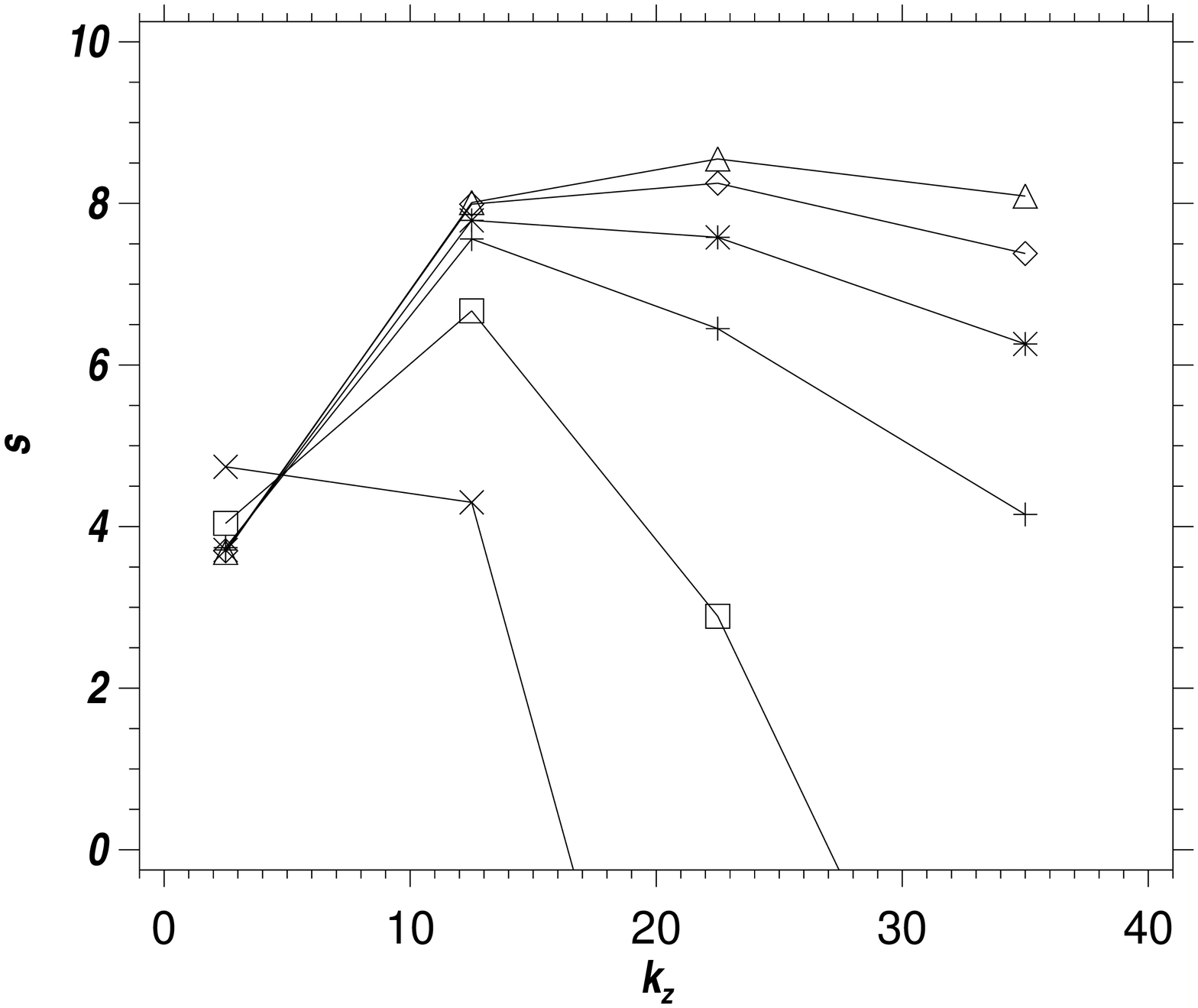}
\includegraphics[width=3in,height=3in]{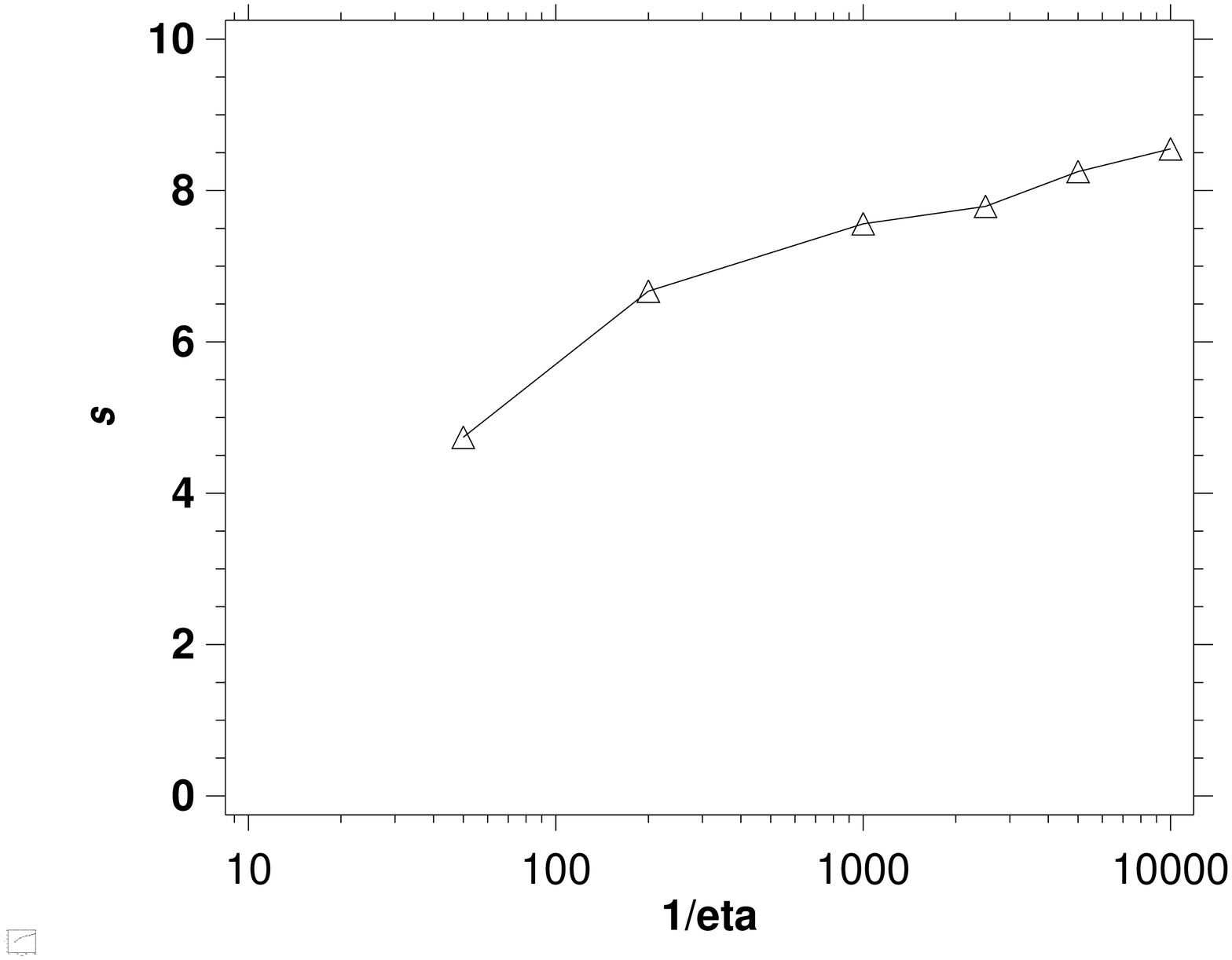}
\caption{Typical and optimal growth-rate curves. Growth-rate as a function of $k_z$ for $\eta=0.02$ (crosses), $0.005$ (squares) $0.001$ (pluses) $0.0005$ (asterisks) $0.0002$ (diamonds) and $0.0001$ (triangles). (b) Optimum growth-rate as a function of $\eta$.Here $V_0=5$ and
$\tau_0=2.0$}
\label{figure9}
\end{figure}

\begin{figure}
\centerline{\includegraphics[width=5in,height=3in]{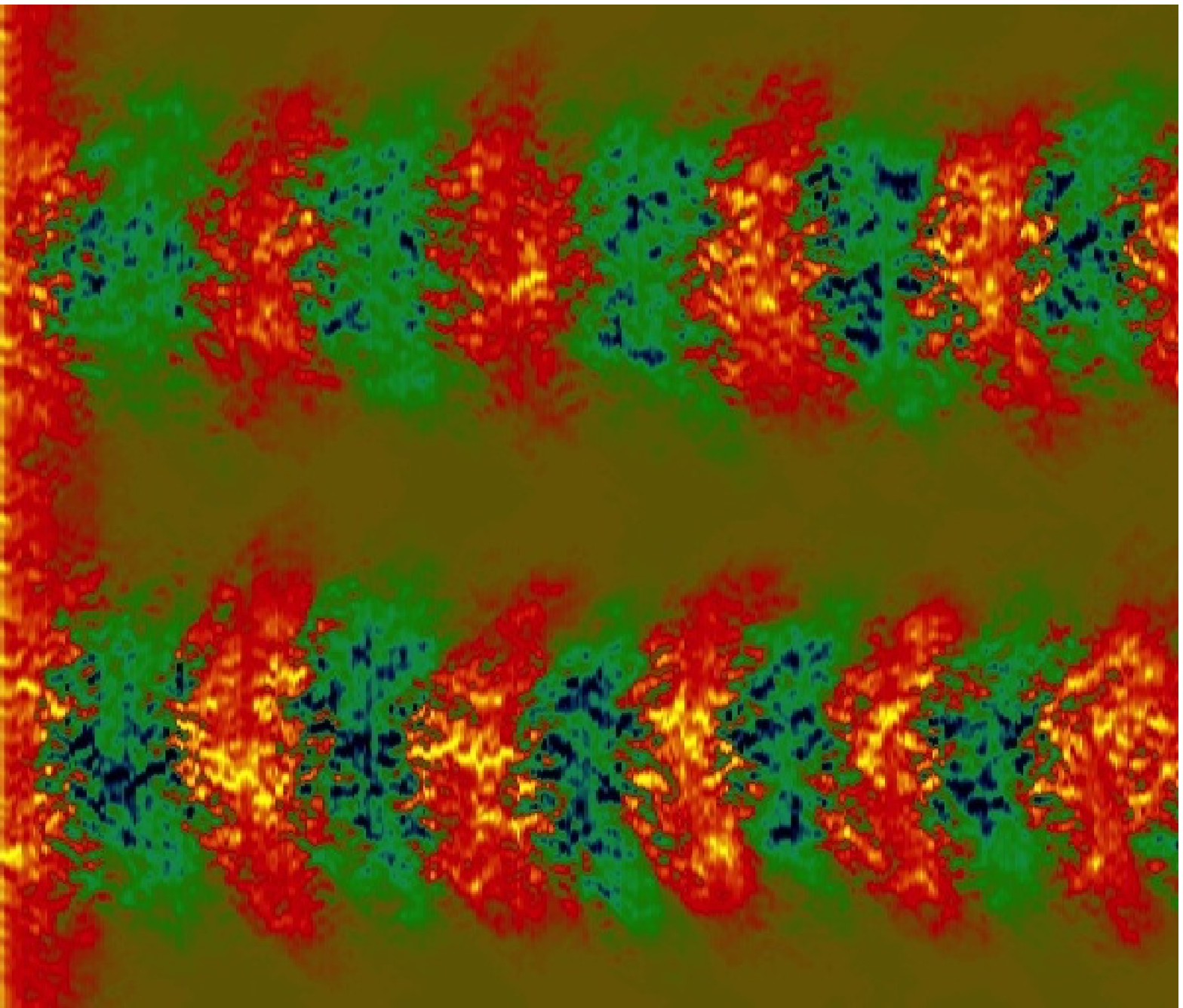}}\hfill\break
\centerline{\includegraphics[width=5in,height=3in]{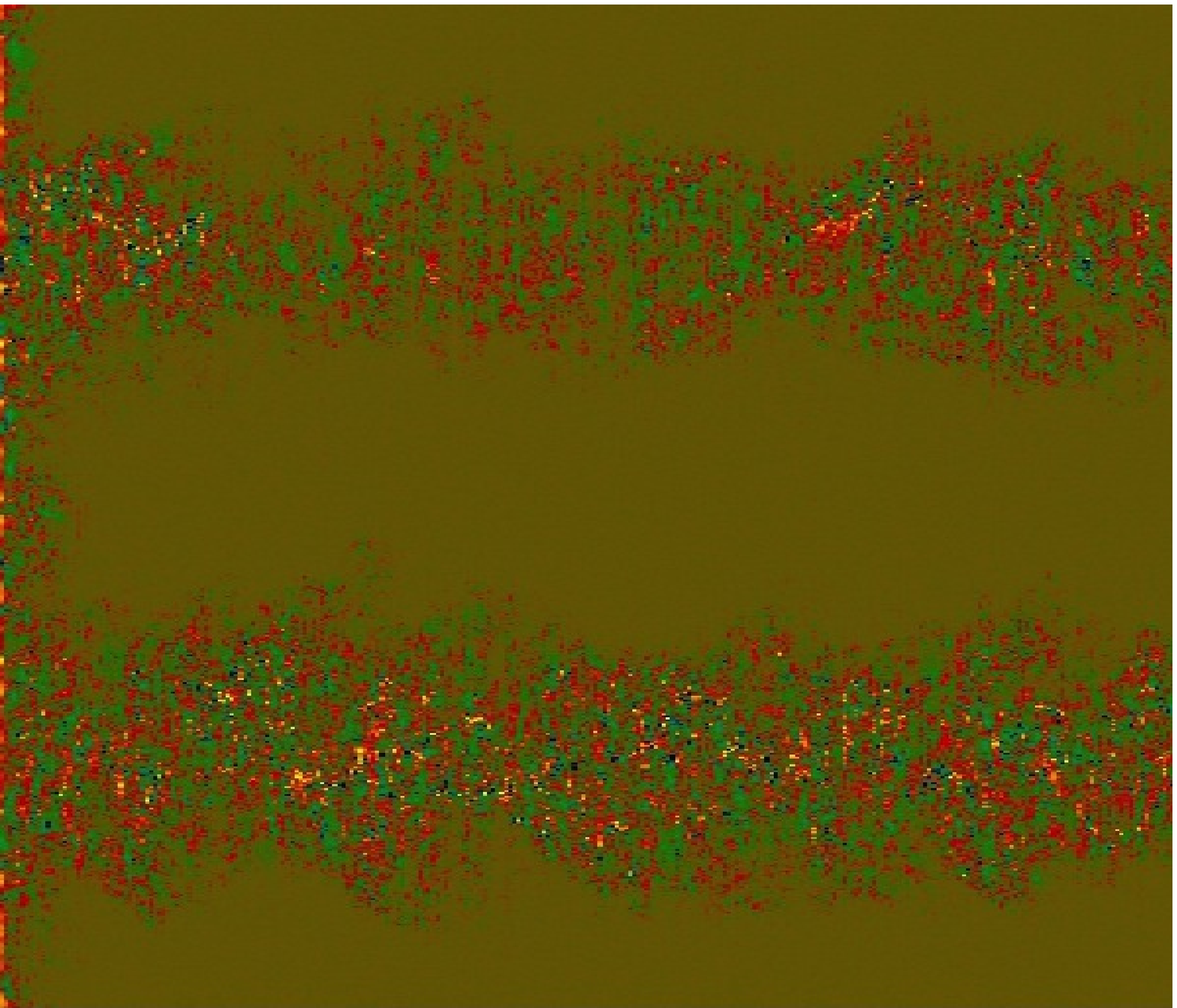}}
\vskip 0.05in
\caption{As for figure~\ref{figure4}, but here $k_{min}=4$, $k_{max}=64$. (a) $\eta=0.02$ and the preferred mode is large-scale. (b) $\eta=0.0001$ and 
the preferred mode is small-scale}
\label{figure10}
\end{figure}


%
\begin{acknowledgments}
This work was partially supported by the Science and Technology
  Facilities Council (STFC) and by the National Science Foundation
  sponsored Center for Magnetic Self-Organisation at the University of
  Chicago. Computations were performed on the STFC supported UKMHD
  consortium cluster (DiRAC) at the
  University of Leeds.   
\end{acknowledgments}


\bibliographystyle{apj}
\bibliography{cattobapj14}

\end{document}